\documentclass[wcp]{jmlr}

\usepackage{bbm}
\usepackage{microtype}
\usepackage{graphicx}
\usepackage{booktabs} 
\usepackage{amsmath,amsfonts}
\usepackage{caption}
\usepackage{multirow}
\usepackage{enumitem}
\usepackage{setspace}

\newcommand{\R}{\mathbb{R}}

\usepackage{lineno}

\makeatletter
\let\Ginclude@graphics\@org@Ginclude@graphics 
\makeatother

\jmlrvolume{222}
\jmlryear{2023}
\jmlrworkshop{ACML 2023}

\title[MEMREC]{Mem-Rec: \underline{Mem}ory Efficient \underline{Rec}ommendation System using Alternative Representation}

 \author{\Name{Gopi Krishna Jha}\textsuperscript{1},
 \Name{Anthony Thomas}\textsuperscript{2},
 \Name{Nilesh Jain}\textsuperscript{1},
 \Name{Sameh Gobriel}\textsuperscript{1},
 \Name{Tajana Rosing}\textsuperscript{2},
 \Name{Ravi Iyer}\textsuperscript{1}
 \\
 \addr \textsuperscript{1}Intel Labs, \addr \textsuperscript{2}University of California, San Diego
}

\editors{Berrin Yan{\i}ko\u{g}lu and Wray Buntine} 

\begin{document}

\maketitle

\begin{abstract}
Deep learning-based recommendation systems (e.g., DLRMs) are widely used AI models to provide high-quality personalized recommendations. Training data used for modern recommendation systems commonly includes categorical features taking on tens-of-millions of possible distinct values. These categorical tokens are typically assigned learned vector representations, that are stored in large embedding tables, on the order of 100s of GB. Storing and accessing these tables represent a substantial performance burden.

Our work proposes \textit{MEM-REC}, a novel alternative representation approach for embedding tables. \textit{MEM-REC} leverages Bloom filters and hashing methods to encode categorical features using two cache-friendly embedding tables. The first table contains raw embeddings (i.e. learned vector representation), and the second table, which is much smaller, contains weights to scale these raw embeddings to provide better discriminative capability to each data point. We provide a detailed architecture, design and analysis of \textit{MEM-REC} addressing trade-offs in accuracy and computation requirements. In comparison with state-of-the-art techniques \textit{MEM-REC} can not only maintain the recommendation quality and significantly reduce the memory footprint for commercial scale recommendation models but can also improve the embedding latency.  In particular, based on our results, \textit{MEM-REC} compresses the MLPerf CriteoTB benchmark DLRM model size by $2900\times$ and performs up to $3.4\times$ faster embeddings while achieving the same AUC as that of the full uncompressed model.
\end{abstract}

\section{Introduction}

Personalized recommendation systems are at the heart of a wide range of applications, including online retail, content streaming, search engines, and social media. The accuracy of these models has a significant business and monetary impact \citep{desai2021semantically}, 
The most accurate modern systems are typically based on large deep neural networks, which account for a significant use of computational resources. At Facebook, for example, up to 72\% of data-center AI inference cycles are devoted to recommender systems \citep{ke2019recnmp}. 

Deep learning recommendation models such as Facebook's ``DLRM'' \footnote{{DLRM}: An advanced, open source deep learning recommendation model, URL: https://ai.facebook.com/blog/dlrm-an-advanced-open-source-deep-learning-recommendation-model}, are trained on a mix of both numeric and categorical features. The categorical features can take on an extremely large number of possible values (easily in the hundreds of millions), which makes representing them a burdensome proposition. A typical design assigns each possible feature value one, or more, dense vector representations, called "embedding," which are stored in a large table and tuned during the learning process. 

When the cardinality of the categorical features is high, embedding tables typically require substantial amounts of memory. Consequently, embedding table lookup and data operations are the main bottleneck during training and inference for these models \citep{ke2019recnmp,2020Intel}. The sheer size of these tables renders traditional acceleration techniques like prefetching, dataflow, and caching completely ineffective \citep{rm2}. Furthermore, embedding operations have orders of magnitude lower compute intensity (Flops/Byte) than fully connected layers \citep{ke2019recnmp}. 

 As a result, there has been a substantial amount of research attention on improving the efficiency of embedding tables. One line of work has focused on compressing embedding tables using techniques like matrix factorization and hashing~\citep{compositional_embedding_compression,mixed_embedding_compression,desai2021semantically, learn21, mixed_dimension}. An alternative approach is based on constructing space-efficient representations of the input data itself. The basic idea is to map a sparse categorical input $x \in \{0,1\}^{m}$, where $m$ is the size of the categorical alphabet, to a new representation $\phi(x) \in \{0,1\}^{d}$, where $d$ is (hopefully) much less than $m$, thereby allowing one to reduce the number of embeddings that need to be stored. We here investigate a simple, yet remarkably powerful, technique for constructing the map $\phi$ based on a set of Bloom filters.

While prior work has also considered similar architectures for this purpose \citep{bloom_encoding}, these approaches have been unable to maintain an acceptably high level of predictive accuracy while simultaneously providing low memory use \citep{embedding_learning}. Our work proposes a new technique, called MEM-REC, that uses a dual encoding process for categorical inputs. The first encoder represents a high-dimensional categorical input as a low-dimensional binary string using a Bloom filter. To mitigate potential loss in accuracy from hash-collisions in the Bloom filter, we introduce a second encoder that re-scales the embeddings in a data-dependent fashion. Our dual encoders can construct embedding representations efficiently ``on-the-fly'' using hashing without the need to maintain any explicit mapping between symbols in the alphabet and offsets in an embedding table.   

In this work, focusing on the widely-used DLRM model for recommender systems,  we evaluate MEM-REC architecture as an efficient alternative embedding table representation. Our  technical contributions are as follows:
\begin{itemize}[noitemsep]
        \item We present a novel algorithm for encoding large-scale embedding tables using a dual Bloom encoding approach to generate memory-efficient embeddings.
        \item We show that even under relatively strict constraints on memory size, MEM-REC can maintain or improve the recommendation quality of commercial-scale DLRMs.
        \item Moreover, we show that MEM-REC can achieve better model size reduction than the state-of-the-art techniques and the compressed embedding tables can fit in the last level caches (LLC) of modern server class hardware. In particular, \textit{MEM-REC} reduces the MLPerf Criteo TeraByte DLRM model size by $2900\times$ with no loss in AUC.
        \item We provide detailed performance characterization of MEM-REC's embedding on a server class CPU system. We show that MEM-REC can alleviate the memory bandwidth bottleneck of the embedding workload and can perform up to $3.4\times$ faster embeddings than the DLRM baseline.
\end{itemize}

The rest of this paper is organized as follows; Section ~\ref{sec:related_work} covers background on DLRM and related work. In Section ~\ref{sec:encoding}, we describe our categorical feature encoding approach. The architecture details of the proposed model are covered in Section ~\ref{sec:bloomed_model}. In Section ~\ref{sec:results}, we present an experimental evaluation of our model.  We conclude the paper in Section ~\ref{sec:future_work} and highlight possible future work. 


\section{Background and Related Work}
\label{sec:related_work}


\subsection{DLRM} \label{sec:prelims}

\begin{figure}
\centering
\begin{minipage}{.48\linewidth}
  \centering
  \includegraphics[width=\linewidth]{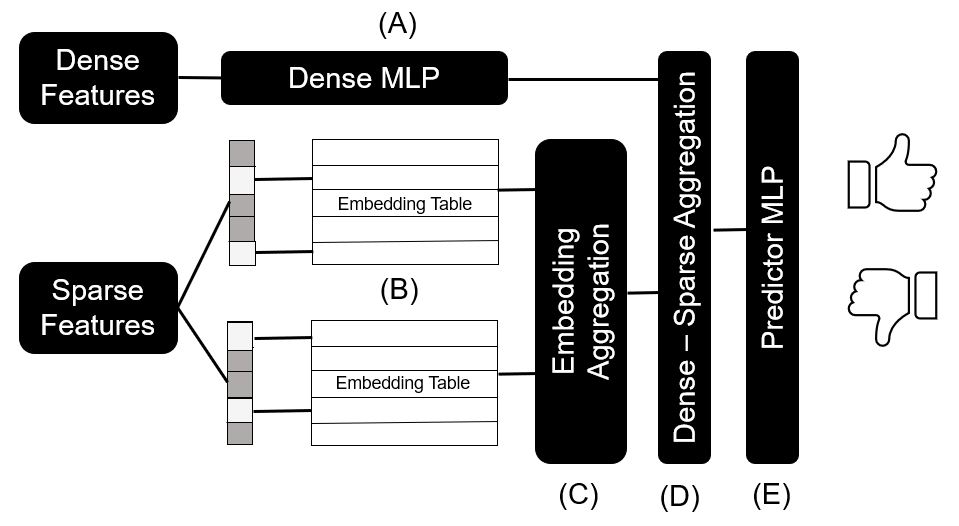}
  \captionof{figure}{DLRM Recommendation Pipeline. An overwhelming majority of the trainable parameters in DLRM come from embedding tables.}
  \label{fig:dlrm}
\end{minipage}%
\hfill
\begin{minipage}{.48\linewidth}
  \centering
  \includegraphics[width=\linewidth]{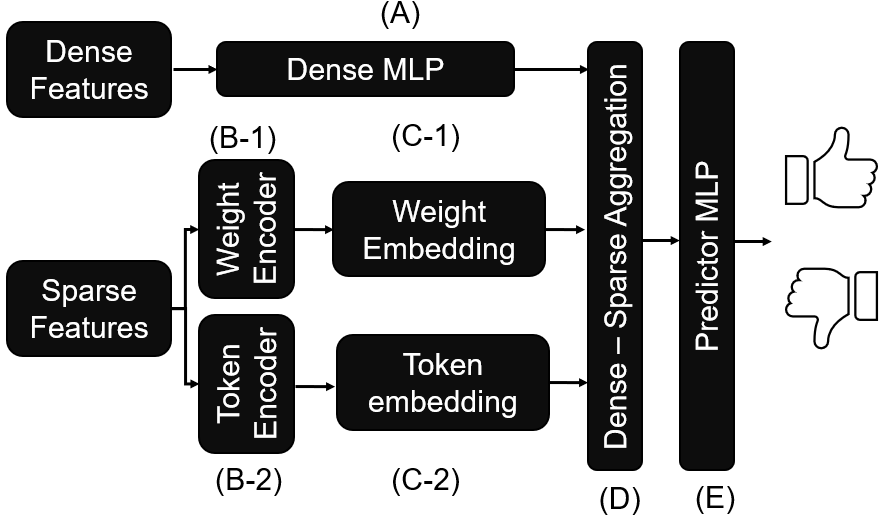}
  \captionof{figure}{Architecture of the MEM-REC Model. Irrespective of the number of categorical features, the MEM-REC model creates only two embedding tables with size scaling just logarithmically in the alphabet size.}
  \label{fig:hybrid_arch}
\end{minipage}
\end{figure}

We focus our approach around the DLRM model developed by Facebook \citep{naumov2019deep}, and summarized in Figure ~\ref{fig:dlrm}, which is the de-facto industry standard, and the reference benchmark for recommendation systems \citep{mlperf_inference, 9407135}. 
DLRM is structurally similar to other state-of-the-art models. As noted above, recommender systems typically rely on a mix of numeric and categorical features, which must be embedded into a common vector representation that can be ingested by a deep neural network. Like most architectures, DLRM maintains two separate encoding pipelines, which are subsequently merged to generate a composite representation of both feature types. 

The entire architecture (including the embedding tables) is trained end-to-end using back-propagation. The embedding tables contain an overwhelming majority (more than 99\% for the Criteo Kaggle \footnote{Criteo kaggle dataset, URL: https://www.kaggle.com/c/criteo-display-ad-challenge}, Avazu \footnote{Avazu mobile advertisement dataset, URL: https://www.kaggle.com/competitions/avazu-ctr-prediction/data} and Criteo Terabyte\footnote{Criteo 1tb click logs dataset, URL: https://ailab.criteo.com/download-criteo-1tb-click-logs-dataset} datasets) of the trainable parameters in the model. Thus, reducing the number of trainable parameters in the embedding tables represents a high-impact area to extract efficiency gains, and is the primary focus of our architecture.

\subsection{Related Work}

One way to mitigate the inefficiencies caused by the memory-heavy embedding tables is to use hardware accelerators. The primary consideration in designing such an accelerator is to reduce data movement between the memory and the compute units. Works like \cite{ke2019recnmp} use near-memory-processing to solve the memory bandwidth bottleneck by bringing compute close to the memory. While these hardware solutions improve efficiency and alleviate the memory bandwidth to some extent, they still suffer from the high storage space required for storing these large embedding tables, the cost of specialized hardware, and less flexibility in comparison to algorithmic solutions.

Algorithmic techniques to reduce the size of embedding tables in DLRM have received substantial research attention in recent years. Broadly speaking, we can divide these works into two categories: (1) embedding compression, and (2) representation compression.

\subsubsection{Embedding Compression}

Embedding compression techniques approach the problem by reducing the storage requirements of the embedding tables. The simplest such approach in this category is to use the standard ``hashing trick,'' which maps embeddings from the full embedding table into a smaller ``compressed'' embedding table \citep{hashing_trick}. The disadvantage of this approach is that multiple ``full'' embeddings, because of hash collisions, are mapped to the same compressed embedding, and hence, some tokens are indistinguishable which degrades the model accuracy. To address this limitation, \cite{compositional_embedding_compression} proposed to partition the embedding table using complementary partitions, and then apply compositional operators on vectors read from each partitioned table to produce the final embeddings. Work in \cite{ttrec} extends this idea and used tensorization techniques to decompose large embedding tables lookups into a sequence of matrix products of reduced dimensions, thus trading off memory capacity and bandwidth with computation. Another work, \cite{mixed_embedding_compression} exploits the observation that frequencies of categorical values are often skewed, and therefore, adapt the embedding dimensions according to the frequency of categorical values. \cite{desai2021semantically} uses a locality sensitive hashing (LSH) function instead of random hashing so semantically-related embeddings (commonly appearing together in the data) are more likely to share a particular dense representation, and hence improve on the accuracy degradation due to hash collisions.  

In this category, one of the recent works ROBE~\citep{robe} extended the basic idea of \cite{hashing_trick} by hashing chunks/blocks of the embeddings (instead of individual tokens), this leads to an improved overhead and better accuracy which allowed for large compression ratio when compared to \cite{hashing_trick}. However, ROBE suffers from the drawback that once a block size is defined then ROBE's embedding table grows linearly with the number of blocks (i.e. proportional linearly with the dataset size); hence, a block size dictates a fixed compression ratio irrespective of the data. As a result, this makes fitting a commercial-scale embedding table in a server class hardware last level caches (LLC) difficult. \textit{MEM-REC}, differs by using a two-stage architecture, the first stage, similarly uses hashing, but instead uses multiple hash functions to define the mapping between categorical tokens and a set of indices in a small Bloom filter data structure (i.e. a token signature). While the second stage (weight encoder) is a much smaller data structure that allows the model to distinguish between tokens that share the embedding in the first table (i.e. tokens with partial overlap in their corresponding signatures), and hence improves the model accuracy while achieving a larger compression ratio as we discuss in details in Section~\ref{sec:results}. The advantage of \textit{MEM-REC} is that model size grows only logarithmically with the data-set alphabet size, and hence, provides much better scalability for commercial-scale recommendation embedding tables. Furthermore, \textit{MEM-REC}'s dual encoder architecture allows the system designer to exploit the advantage of fitting each encoder at different levels of cache hierarchy in a server platform (e.g. fit token encoder in LLC and fit weight encoder in L2 cache; as we discuss later).

\subsubsection{Representation Compression}

Representation compression techniques reduce the number of embeddings to store by leveraging more parsimonious representations for categorical data. 

Recent work in \cite{embedding_learning} use a hashing based scheme to map categorical data onto a dense vector in $\mathbb{R}^{d}$ for $d \ll m$. These dense representations can then be consumed by an MLP-style neural network which produces the final embedding. This work differs significantly from ours in that the categorical representations are \emph{dense} (e.g. most components are non-zero) and real-valued. The advantage of real-valued representations is that they do not suffer from errors related to collisions. However, these embeddings are fairly large in practice - $d = 1024$ in the context of \cite{embedding_learning} -- and the MLP used to process them must be implemented using a dense matrix-vector multiplication. By contrast, in our approach based on Bloom filters, the MLP used to construct the final embedding can be implemented by computing an element-wise sum of $\approx 10$ vectors. 

Earlier work in \cite{bloom_encoding} considered using Bloom filters to encode high-dimensional categorical data. However, authors of ~\cite{embedding_learning} have shown that at small memory footprints such technique is unable to maintain the required recommendation quality, and hence, is only limited to simple AI architectures and does not scale for commercial scale DLRM model. MEmCom technique presented in \cite{apple-mlsys23} has also explored two-stage architecture, however, the design of the two stages differs significantly from MEM-REC. MEmCom uses hashing trick~\citep{hashing_trick} in the first stage followed by a one-to-one token-to-weight mapper in the second stage. The second stage contains one trainable weight per token, hence suffers with table size explosion, which leads to a very small compression (up to 40x). Furthermore, it suffers from a relatively severe degradation in accuracy ($> 1\%$) and hence is not applicable to commercial scale recommendation models as they are highly sensitive to even small losses in accuracy~\citep{desai2021semantically}.

\section{The MEM-REC Model} \label{sec:bloomed_model}


In the following section, we describe our model architecture. Like DLRM, MEM-REC follows a three-stage (feature encoding, feature aggregation and prediction) process. To encode numeric data, we use DLRM's default MLP style dense network (Figure \ref{fig:hybrid_arch}: item A). Our main innovation, which we now describe, is in the design of the categorical encoding.

\subsection{Categorical Features} \label{sec:encoding}

\begin{figure*}[t]
\centering
     \includegraphics[width=0.8\textwidth]{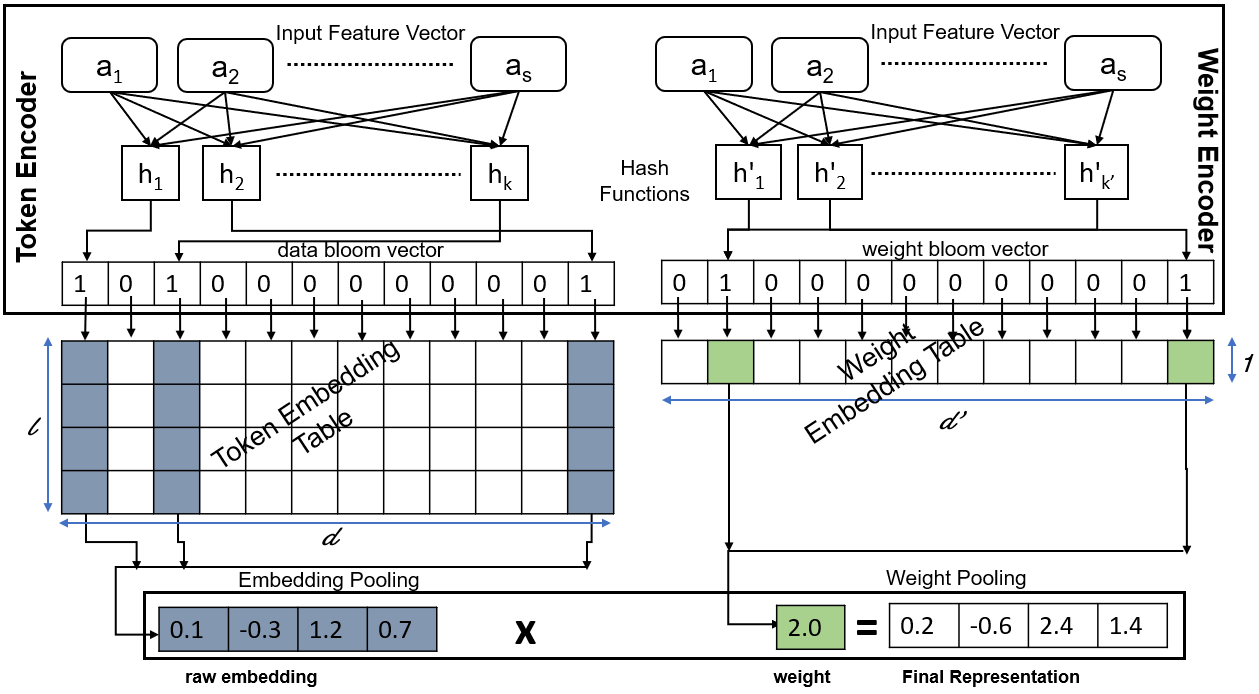}
     \caption{Sparse Feature Encoding Flow in MEM-REC. The token encoder generates raw embeddings and the weight encoder changes the scale of the token encoding to mitigate the effect of hash-collisions}
     \label{fig:encoding_flow}
\end{figure*}

Let $\mathcal{A} = \{a_{1},...,a_{m}\}$ denote the alphabet from which categorical data is drawn, $x$ be a vector of $s$ categorical features to be encoded, and observe that we may think of $x$ as a set drawn from $\mathcal{A}$. In DLRM, $x$ is represented as a ``multi-hot'' encoding $\delta(x)$, that contains exactly $s$ ones, whose locations encode the symbols in $x$ (e.g. $\delta(x)$ is the characteristic vector of the set $x$). The embedding operation can then be written as $z(x) = M\delta(x)$, where $M \in \R^{l \times m}$ is the embedding table, and $l$ is the embedding dimension. Our approach is to compress $\delta(x)$ to a $d \ll m$ dimensional bitvector using dual encoders, which allows us to reduce the number of columns in $M$ to $d$.

\subsection{Categorical Feature Encoder} \label{sec:bloom_encoding}

Let $h_{1}, ... ,h_{k}$ be a set of $k$ independent hash functions that take a symbol $a \in \mathcal{A}$ as input and return an index between $0,...,d-1$ as output. To embed a symbol $a \in \mathcal{A}$, we initialize $\psi(a) = 0_{d}$ to the zeros vector. Then, as shown in Figure ~\ref{fig:encoding_flow}, the token encoder computes $h_{1}(a),...,h_{k}(a)$ and sets the resulting indices to one. That is: 
\[
    \psi(a)^{(j)} = 
        \begin{cases}
            1 & \text{ if }h_{i}(a)=j\text{ for any }i \in [k]\\
            0 & \text{otherwise}
        \end{cases}
\]
The encodings of tokens in a feature vector $x \subset \mathcal{A}$, are then pooled using the rule:
\[
    \phi(x) = \max_{a \in x} \psi(a) 
\]
where the maximum is applied element-wise. This is exactly the encoding rule of the Bloom filter, a canonical approximate data structure used to represent sets \citep{bloom_basic}. As before, we construct the representation of $x$ by pooling together the columns in $M$ corresponding to ``active'' elements in $\phi(x)$. That is, via $z(x) = M\phi(x)$.

In the ideal case that there are no hash-collisions, then $\phi(x)$ is again simply a unique representation of the input $a$ and can be interpreted in the same manner as the ``multi-hot'' encodings used in DLRM. However, when $d \ll m$, which is necessary to achieve memory savings, collisions are unavoidable, which means that different symbols will sometimes be forced to share embeddings (e.g. columns in $M$). This may be undesirable because it can induce correlation between the representations of completely unrelated inputs. While the probability of collision can be made small by choosing $d$ to be large relative to $k$, this detracts from the original goal of making $M$ (i.e. memory footprint) as small as possible. 

To keep $M$ small, while mitigating the effect of hash-collisions, we introduce a second stage of encoding, that we term ``weight-encoding,'' which multiplies $\phi(x)$ by a weight $\alpha(x)$, that depends on the input feature vector. This means that, even if two embeddings point in nearly the same direction, their magnitudes are scaled by different factors, making them easier to differentiate. 

\subsection{Weight Encoder}

As shown in Figure \ref{fig:encoding_flow}, the weight encoder is similar in structure to the token encoder and is also implemented using a Bloom filter, which is instantiated using a fresh set of $k'$ hash-functions. The output of the weight-Bloom-filter is another bit-vector $\phi'(x) \in \{0,1\}^{d'}$, which is used to compute the weight $\alpha(x) = \phi'(x)^{T}w$, where $w \in \R^{d'}$ is a vector of trainable weights. The final embedding of an input $x$ is then given by $z(x) = \alpha(x)M\phi(x)$, where $\alpha(x)$ is a scalar, $M \in \R^{l \times d}$, and $\phi(x) \in \{0,1\}^{d}$. To reiterate: the ``weight-encoding'' allows us to change the scale of the token encoding, thereby mitigating the effect of hash-collisions in the token encoder. From a practical standpoint, this allows us to hash onto a smaller range in the token encoder (e.g. use a smaller $d$), which means we need to store fewer embeddings (a smaller number of columns in $M$), which translates into increased memory savings. It also allows us to reduce $k$ (the number of hash-functions in the token encoder), thereby reducing the number of memory accesses to the token embedding table, at the cost of more frequent accesses to the weight embedding table, which is much smaller and can fit in lower levels of cache. As shown in Figure \ref{fig:we_impact}, using the weight encoder not only reduces the number of unresolved collisions but also achieves faster overall embedding latency than the one-stage token-encoder-only model. 


\begin{figure}[htbp]
\floatconts
{fig:we_impact}
{\caption{Effect of weight encoder on collisions, and embedding latency of a $50000\times128$ size Criteo-TB MEMREC Model. Weight encoder reduces the number of unresolved collisions and helps reduce the memory access latency by prioritizing frequent accesses to the feather-light weight embedding table which fits in the L2 cache.}}
{
\subfigure{
\label{fig:we_collisions}
\includegraphics[width=0.45\linewidth]{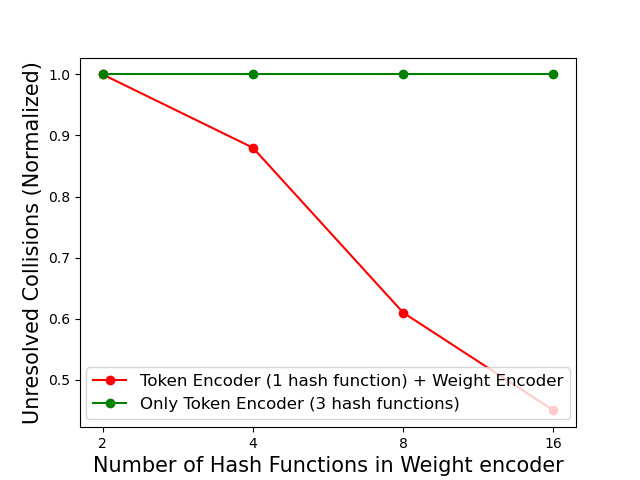}
}\qquad
\subfigure{
\label{fig:we_latency}
\includegraphics[width=0.45\linewidth]{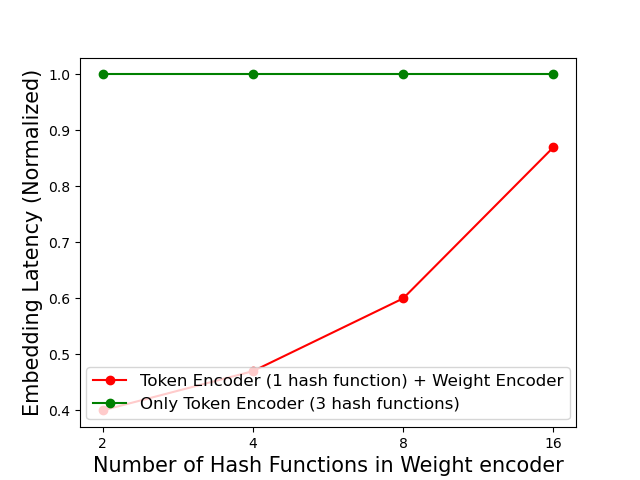}
}
}
\end{figure}

\subsection{Architecture Summary}

To summarize the discussion above: to encode the categorical data, MEM-REC simultaneously passes inputs to the token encoder and the weight encoder as explained above. For each data point, the token encoder (Figure \ref{fig:hybrid_arch}: item B2) and the weight encoder (Figure \ref{fig:hybrid_arch}: item B1) generate one sparse binary vector each of size $d$ and $d^{'}$ respectively.

The output of the token encoder is passed to an embedding table (Figure \ref{fig:hybrid_arch}: item C2) of size $l \times d$, where $l$ represents the dimension of each embedding vector. \textit{MEM-REC} selects columns from this embedding table whose corresponding bits are set in the binary Bloom vector and pools them together to get the raw embedding for the data point. The output of the weight encoder is passed to a much smaller embedding table (Figure \ref{fig:hybrid_arch}: item C1) of size $d^{'} \times 1$. \textit{MEM-REC} generates a weight for each data point by pooling the indices of this embedding table as indicated by 1s in the corresponding Bloom vector. The final embedding is computed by multiplying the raw embedding vector with the generated weight. This process is explained in Figure \ref{fig:encoding_flow}.
The key distinction from the standard DLRM architecture is that our approach only requires storing $d$ embeddings, as opposed to $m$, where $d \ll m$ by several orders of magnitude. The output of the embedding tables used for encoding categorical features is then merged with the output of the dense MLP (Figure \ref{fig:hybrid_arch}: (D)) and the result is passed to a predictor MLP which outputs the prediction (Figure \ref{fig:hybrid_arch}: item E).

\section{Experiments} \label{sec:results}

We organize our experiments to answer the following questions:
\begin{itemize}[noitemsep]
    \item \textbf{RQ1}: How do the settings of key MEM-REC parameters like the number of hash-functions ($k$), and the hash-alphabet size ($d$), affect memory use and recommendation quality?
    \item \textbf{RQ2}: For a given memory budget, how does MEM-REC's recommendation quality compare against the traditional and recent embedding compression approaches?
    \item \textbf{RQ3}: To what extent can MEM-REC compress a model without compromising recommendation quality? Can these compressed models fit in the last level caches of modern server CPUs?
    \item \textbf{RQ4}: What are the performance implications of MEM-REC on modern server-class CPUs?
\end{itemize}

\subsection{Experimental Setup}
    \subsubsection{Datasets}
    We use Avazu, Criteo-Kaggle and Criteo-Terabyte, three large popular open source benchmark CTR datasets to evaluate MEM-REC. Due to space limitations, we provide dataset details and partitioning methodology in the appendices.

    To keep runtimes tractable when comparing a large number of hyper-parameters, for RQ1 (Section ~\ref{sec:rq1}) and RQ2 (Section ~\ref{sec:rq2}), we focus on the Criteo-Kaggle and Avazu datasets. We compare our architecture against DLRM and other state-of-the-art techniques on the full Criteo-TB dataset in Section~\ref{sec:model_compression}."

    \subsubsection{Basic Setup}
    We used the DLRM PyTorch implementation as publicly provided on GitHub.\footnote{{DLRM} Github, URL: https://github.com/facebookresearch/dlrm}
    
    \textbf{DLRM:}
    We follow the basic setup like optimizer, sparse feature size, MLP architecture as suggested by the DLRM paper \citep{naumov2019deep} and their GitHub implementation. 
    These parameters are summarized in the appendices.

    \textbf{MEM-REC Model:}
    We follow the same setup as DLRM for the common parts (items (A), (D) and (E) in Figures ~\ref{fig:dlrm} and ~\ref{fig:hybrid_arch}) of the recommendation pipeline. For the new stages (items (B) and (C) in Figure ~\ref{fig:hybrid_arch}) however, we perform additional tuning to select parameters like the number of hash functions ($k$ and $k^{'}$), the number of bits in the two Bloom filters ($d$ and $d^{'}$) and the size of raw embedding vector ($l$) in order to find different configurations of MEM-REC to suit different AUC and efficiency requirements.
    
    \subsubsection{Batch Size and Number of Epochs}
    
    \textbf{Criteo-Kaggle and Avazu:} The original DLRM paper \citep{naumov2019deep} uses a batch size of 128 and runs the model just for 1 epoch. However, subsequent work \citep{large_batch_ctr,desai2021semantically} recommends using larger batch sizes for click-through-rate prediction (CTR) tasks. For both DLRM and MEM-REC models, we use the batch size of 1024 and we run the training for 4 epochs. We observe that increasing batch size and epoch from 128 to 1024 and 1 to 4 respectively leads to slightly better AUC for both DLRM and MEM-REC models.
    
    \textbf{Criteo-Terabyte:} We use the batch size of 2048 as used by mlperf DLRM implementation.\footnote{{MLPerf} Inference, URL: https://mlcommons.org/en/inference-datacenter-11} However, the caveat while using MEM-REC on Criteo-Terabyte is that the model requires 6 epochs to converge as against 1 epoch used by the baseline DLRM implementation. We have achieved this without extensive hyper-parameter tuning and leave this optimization for future work.

\subsection{Accuracy and Efficiency implications of MEM-REC parameters (RQ1)} \label{sec:rq1}
    In DLRM, embedding tables contribute the majority of the model parameters. For instance, for all three datasets considered here, over 99\% of the parameters in DLRM are in the embedding tables. MEM-REC uses two Bloom filters to efficiently represent categorical data which dramatically reduces the number of parameters to store. In this section, we summarize the new parameters and discuss their impact on accuracy, model size and efficiency.

        \begin{figure}[htb]
        \floatconts
        {fig:k_sweep}
        {\caption{AUC for different values of $k$ at $k^{'}=2$}}
        {
        \subfigure[Small Model: \textit{(d,l)=(10000,16)}]{
        \label{fig:k_sweep_small_model}
         \includegraphics[width=0.45\linewidth]{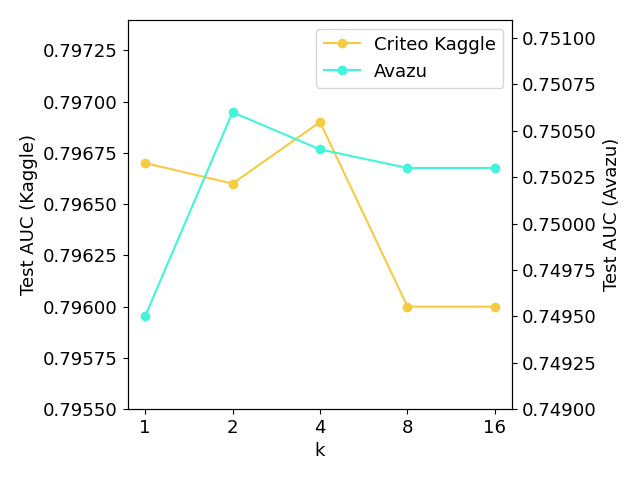}
        }\qquad
        \subfigure[Large Model: \textit{(d,l)=(150000,512)}]{
        \label{fig:k_sweep_large_model}
        \includegraphics[width=0.45\linewidth]{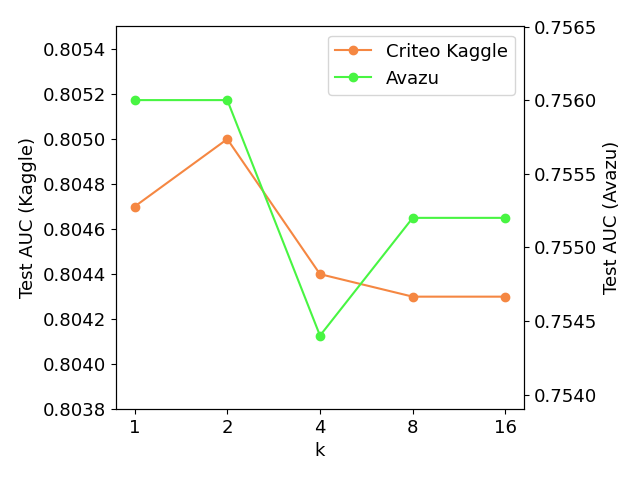}
        }
        }
        \end{figure}

        \begin{figure}[htb]
        \floatconts
        {fig:kdash_sweep}
        {\caption{AUC for different values of $k^{'}$ at $k=1$. To produce optimal recommendation quality, the weight encoder typically uses about $2-8$ hash functions for the small models and $2-4$ hash functions for the larger models.}}
        {
        \subfigure[Small Model: \textit{(d,l)=(10000,16)}]{
        \label{fig:kdash_sweep_small_model}
         \includegraphics[width=0.45\linewidth]{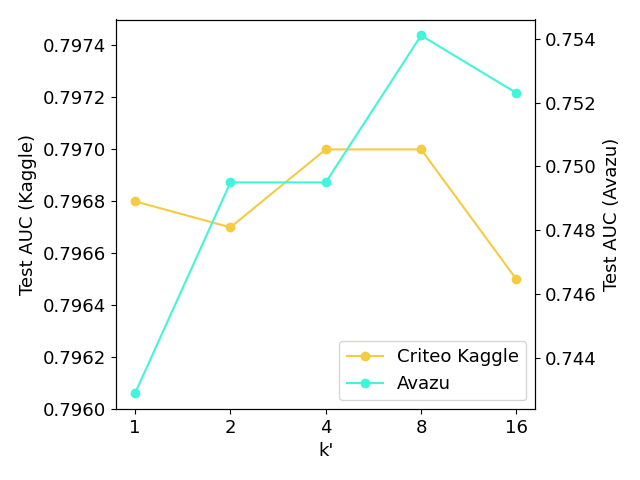}
        }\qquad
        \subfigure[Large Model: \textit{(d,l)=(150000,512)}]{
        \label{fig:kdash_sweep_large_model}
        \includegraphics[width=0.45\linewidth]{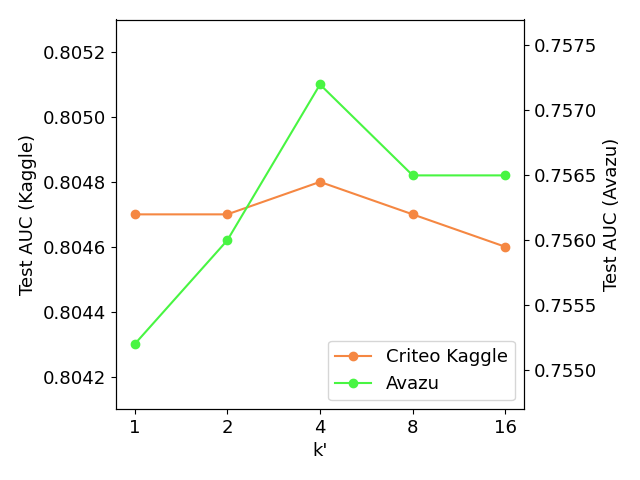}
        }
        }
        \end{figure}

        \begin{figure}[htb]
        \floatconts
        {fig:dl_sweep}
        {\caption{AUC gain vs Bloom-vector length (d) and embedding vector length (l). Increasing $d$ and $l$ improve AUC up to a point beyond which the models start overfitting.}}
        {
        \subfigure[AUC vs d for small \& large embedding vectors]{
        \label{fig:d_sweep}
         \includegraphics[width=0.45\linewidth]{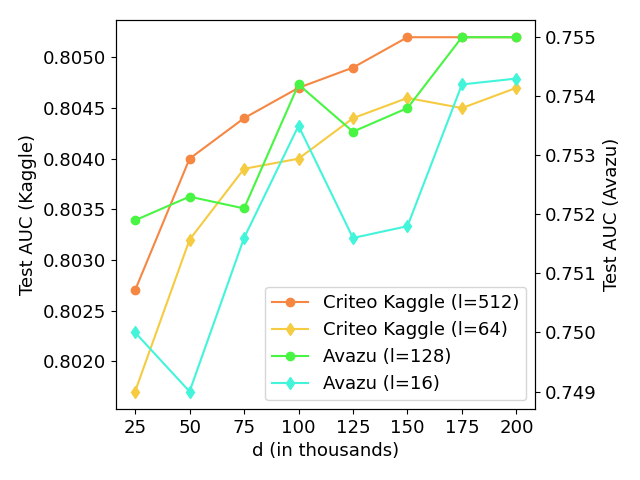}
        }\qquad
        \subfigure[AUC gain vs l for large (d=125,000) \& small (d=10,000) models]{
        \label{fig:l_sweep}
        \includegraphics[width=0.45\linewidth]{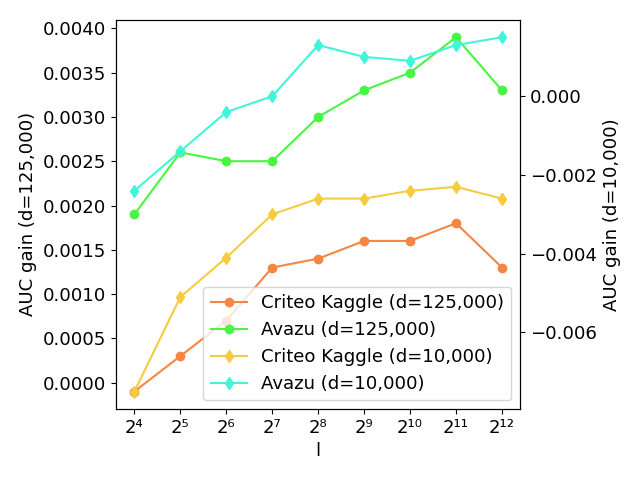}
        }
        }
        \end{figure}

        \subsubsection{Number of hash functions ($k$ and $k^{'}$)}
        The token encoder and the weight encoder use $k$ and $k^{'}$ independent hash functions respectively to encode the categorical data. These values determine how many vectors the model has to access and pool in the corresponding compressed embedding table for every data point and hence they determine the quality of binary encoding and impact the accuracy. The values of $k$ and $k^{'}$ have no impact on the model size. Figures ~\ref{fig:k_sweep} and ~\ref{fig:kdash_sweep} show the AUC of the model for different values of $k$ and $k^{'}$ respectively. For the token encoder, we find that while using about $2-4$ hash functions yields the best results for small models, large models can manage to reach optimal AUC with as low as $1-2$ hash functions. The weight encoder typically uses about $2-8$ hash functions for the small models and $2-4$ hash functions for the larger models to produce optimal recommendation quality. 
        
        \textbf{Takeaway 1:} As the weight embedding table can comfortably fit within the L2 cache of modern server CPUs even for very large datasets like the Criteo TB,  MEM-REC can accommodate moderately high $k^{'}$ without significantly impacting the embedding latency.
            
        \textbf{Takeaway 2:} As the token embedding table can often fully fit in last-level caches (LLCs) of modern server systems (Section ~\ref{sec:model_compression}), using small $k$ allows enough room for MEM-REC to take advantage of the latency delta between DRAM and LLC. 
            
        \subsubsection{Embedding table dimension}  
        \begin{itemize}[noitemsep]
        \item \textbf{Length of the binary Bloom vectors ($d$ and $d^{'}$):}
        The token encoder and the weight encoder use the Bloom vectors of size $d$ and $d^{'}$ respectively. These values represent the number of columns in the corresponding compressed embedding table and they impact both accuracy and model-size directly. To keep our design space simple, we use $d^{'} = d$. Model size scales linearly with $d$. As shown in the Figure ~\ref{fig:d_sweep}, AUC is increasing up to around $d = 175,000$ at which point performance starts to saturate - possibly due to overfitting. This is unsurprising since, for a fixed $k$, taking $d$ to be very large is effectively like assigning each symbol a unique embedding, as in the standard implementation of DLRM.
        
        \item \textbf{Length of embedding vector ($l$):}
        This parameter represents the length of the final embedding vector obtained and therefore has a direct impact on the AUC of the model. Figure ~\ref{fig:l_sweep} shows the AUC gained by the models against the DLRM baseline for different embedding vector lengths. For both cases, where the models use large or small Bloom vectors, increasing $l$ helps achieve better AUC up to a point beyond which the model starts to over-fit.
        \end{itemize}

         The resulting compressed embedding table is of size $(l \times d)$ and for every categorical token, the model selects $k$ vectors from this table and pools them to get the raw embedding representation. 

\subsection{Recommendation Quality (RQ2)} \label{sec:rq2}

    \textbf{Metrics}: Consistent with prior work, we use ROC-AUC (Receiver Operating Characteristics-Area Under The Curve) to determine the quality of recommendation generated by our model \citep{robe}. One important point to note here is that in the case of commercial-scale recommender systems, even a 0.001 change in AUC has a significant business impact \citep{desai2021semantically}. 
    
    \textbf{Baselines}: Apart from vanilla uncompressed DLRM, we compare MEM-REC with the following baseline models at different memory budgets. 
    \begin{itemize}[noitemsep]
            \item \textbf{Hashing Trick \citep{hashing_trick}} maps embeddings from the full embedding table onto a smaller ``compressed'' embedding table and is often less effective as it suffers from collision problems.

            \item \textbf{Compositional Embedding (QR Trick) \citep{compositional_embedding_compression}} partitions the embedding table using two complementary partitions for avoiding hashing collision. It applies compositional operators on vectors from each partition to produce the final embedding. 
            \item \textbf{Random Offset Block Embedding Array (ROBE) \citep{robe}} proposes a memory sharing technique which uses universal hash functions on the embedding table chunks to map them in a small circular array of memory.
            \item \textbf{TT-Rec \citep{ttrec}} employs a matrix decomposition approach by decomposing tensor representation of multidimensional data into the product of smaller tensors.
    \end{itemize}

    \textbf{Comparison}: In Figure ~\ref{fig:quality_comparison}, We plot the AUC with different memory budgets (1/16, 1/8, 1/4, 1/2 and 1/1 of the full model size) for Criteo-Kaggle and Avazu datasets. MEM-REC provides comparable or better AUC to ROBE and TT-REC across memory regimes (number of parameters). We discuss the model-size accuracy trade-off in detail in Section ~\ref{sec:model_compression}.

    \begin{figure}[]
    \floatconts
    {fig:quality_comparison}
    {\caption{AUC for different available memory budget. MEM-REC provides comparable or better AUC than SOTA compression techniques across memory regimes.}}
    {
    \subfigure[Criteo-Kaggle]{
    \label{fig:auc_comparison}
    \includegraphics[width=0.45\linewidth]{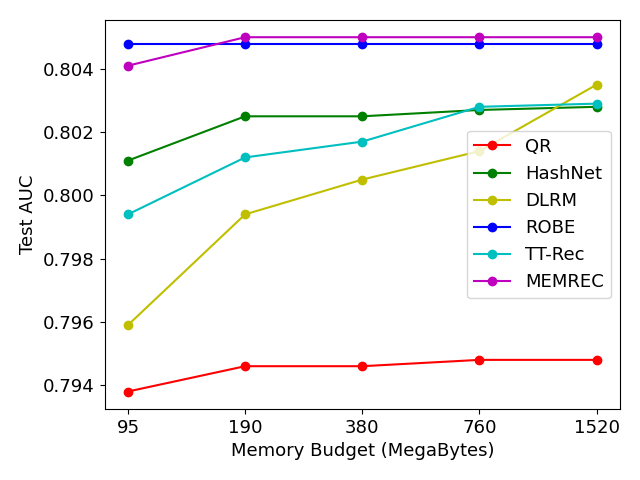}
    }\qquad
    \subfigure[Avazu]{
    \label{fig:accuracy_comparison}
    \includegraphics[width=0.45\linewidth]{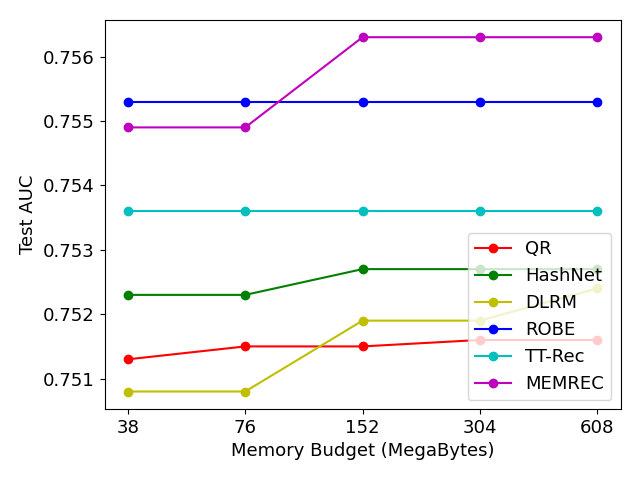}
    }
    }
    \end{figure}
        
\subsection{Model Compression (RQ3)} \label{sec:model_compression}

    We try to answer the question - how far can MEM-REC compress a model without sacrificing the accuracy of the recommendations? Table ~\ref{table:model_size_vs_acc} summarizes the AUC difference and model size reduction factor against baseline DLRM for different model sizes of our approach. We re-emphasize that even a 0.001 loss in AUC is problematic for commercial-scale recommender systems \citep{desai2021semantically}. MEM-REC achieves iso-quality (zero loss in AUC) models with a reduction of $188\times$, $144\times$ and $2904\times$ in model size for Avazu, Criteo-Kaggle and Criteo-Terabyte datasets respectively. As explained in section  ~\ref{sec:encoding}, the MEM-REC approach constructs an approximate representation of a token using a number of bits that scales just logarithmically in the alphabet size. This property allows MEM-REC to gain much higher compression as the size of the dataset grows.
    
    Table~\ref{table:iso_quality_compression} summarizes the maximum compression achieved by MEM-REC, ROBE and TT-REC. While ROBE achieves the best compression for the Avazu and Criteo-Kaggle datasets, MEM-REC outperforms the other methods on the much bigger Criteo-TB dataset. It is important to note that, while smaller models fit into the LLCs with smaller to moderate compression, larger models need more aggressive compression for this to happen. For instance, MEM-REC's iso-quality terabyte model can fully fit into the last level cache (48 MB) of the Intel (Xeon Gold 6338) Icelake family server processor.
    
    \textbf{Takeaway 3:} As MEM-REC constructs embedding tables with size scaling just logarithmically in the alphabet size, the compression factor grows as we move from smaller datasets to larger datasets. This property allows MEM-REC models to fit in LLC even if we are working with a terabyte-scale dataset.

    \begin{table}[]
    
    \scriptsize
    \parbox{.45\linewidth}
    {
            \centering
            \begin{tabular}{|ccc|}
            \hline
            \multicolumn{1}{|c|}{\textbf{\begin{tabular}[c]{@{}c@{}}Parameters\\ (Millions)\end{tabular}}}  & \multicolumn{1}{c|}{\textbf{\begin{tabular}[c]{@{}c@{}}$\Delta$AUC \\ vs DLRM\end{tabular}}} & \textbf{Compression} \\ \hline
            
            \multicolumn{3}{|c|}{\textbf{Criteo-TB}}                                                                                                                                                                                                                                                               \\ \hline
            \multicolumn{1}{|c|}{5}                                                                                                                                        & \multicolumn{1}{c|}{-0.005}                                                          & 4734x                \\ \hline
            \multicolumn{1}{|c|}{8}                                                                                                                                        & \multicolumn{1}{c|}{-0.002}                                                          & 3412x                \\ \hline
            \multicolumn{1}{|c|}{11}                                                                                                                                       & \multicolumn{1}{c|}{0.000}                                                           & 2904x                \\ \hline
            \multicolumn{1}{|c|}{15}                                                                                                                                        & \multicolumn{1}{c|}{0.001}                                                           & 1638x                \\ \hline
            \multicolumn{1}{|c|}{21}                                                                                                                                        & \multicolumn{1}{c|}{0.000}                                                           & 1140x                \\ \hline
            
            \multicolumn{3}{|c|}{\textbf{Criteo-Kaggle}}                                                                                                                                                                                                                                                           \\ \hline
            \multicolumn{1}{|c|}{2}                                                                                                                                        & \multicolumn{1}{c|}{-0.002}                                                          & 251x                 \\ \hline
            \multicolumn{1}{|c|}{4}                                                                                                                                         & \multicolumn{1}{c|}{0.000}                                                               & 144x                 \\ \hline
            \multicolumn{1}{|c|}{5}                                                                                                                                         & \multicolumn{1}{c|}{0.001}                                                           & 101x                 \\ \hline
            \multicolumn{1}{|c|}{7}                                                                                                                                       & \multicolumn{1}{c|}{0.001}                                                           & 78x                  \\ \hline
            \multicolumn{1}{|c|}{10}                                                                                                                                        & \multicolumn{1}{c|}{0.001}                                                           & 53x                  \\ \hline

            \multicolumn{3}{|c|}{\textbf{Avazu}}                                                                                                                                                                                                                                                                   \\ \hline
            \multicolumn{1}{|c|}{0.8}                                                                                                                                         & \multicolumn{1}{c|}{0.000}                                                           & 188x                 \\ \hline
            \multicolumn{1}{|c|}{1.2}                                                                                                                                        & \multicolumn{1}{c|}{0.001}                                                           & 126x                  \\ \hline
            \multicolumn{1}{|c|}{1.6}                                                                                                                                        & \multicolumn{1}{c|}{0.002}                                                           & 95x                  \\ \hline
            \multicolumn{1}{|c|}{2.0}                                                                                                                                       & \multicolumn{1}{c|}{0.002}                                                           & 76x                  \\ \hline
            \multicolumn{1}{|c|}{2.4}                                                                                                                                        & \multicolumn{1}{c|}{0.002}                                                           & 63x                  \\ \hline
            
            \end{tabular}
    \caption{Recommendation quality vs model size}
    \label{table:model_size_vs_acc}
    }
    \hfill
    \parbox{.45\linewidth}
    {
            \centering
            \begin{tabular}{|ccc|}
            \hline
            \multicolumn{1}{|c|}{\textbf{Technique}} & \multicolumn{1}{c|}{\textbf{\begin{tabular}[c]{@{}c@{}}Compression\\  (iso-quality)\end{tabular}}} & \textbf{\begin{tabular}[c]{@{}c@{}}Can \\ fit in a \\ 48MB \\ L3\\  cache\end{tabular}} \\ \hline
            
            \multicolumn{3}{|c|}{\textbf{Criteo-TB}}                                                                                                                                                                                       \\ \hline
            \multicolumn{1}{|c|}{ROBE}               & \multicolumn{1}{c|}{1000x}                                                                         & \text{\sffamily X}                                                                             \\ \hline
            \multicolumn{1}{|c|}{TT-REC}             & \multicolumn{1}{c|}{112x}                                                                          & \text{\sffamily X}                                                                             \\ \hline
            \multicolumn{1}{|c|}{\textbf{MEM-REC}}           & \multicolumn{1}{c|}{\textbf{2904x}}                                                                         & \checkmark                                                                            \\ \hline
            
            \multicolumn{3}{|c|}{\textbf{Criteo-Kaggle}}                                                                                                                                                                                   \\ \hline
            \multicolumn{1}{|c|}{ROBE}               & \multicolumn{1}{c|}{1000x}                                                                         & \checkmark                                                                            \\ \hline
            \multicolumn{1}{|c|}{TT-REC}             & \multicolumn{1}{c|}{117x}                                                                          & \checkmark                                                                            \\ \hline
            \multicolumn{1}{|c|}{\textbf{MEM-REC}}           & \multicolumn{1}{c|}{\textbf{144x}}                                                                          & \checkmark                                                                            \\ \hline
            
            \multicolumn{3}{|c|}{\textbf{Avazu}}                                                                                                                                                                                           \\ \hline
            \multicolumn{1}{|c|}{ROBE}               & \multicolumn{1}{c|}{1000x}                                                                         & \checkmark                                                                            \\ \hline
            \multicolumn{1}{|c|}{TT-REC}             & \multicolumn{1}{c|}{143x}                                                                          & \checkmark                                                                            \\ \hline
            \multicolumn{1}{|c|}{\textbf{MEM-REC}}           & \multicolumn{1}{c|}{\textbf{188x}}                                                                          & \checkmark                                                                            \\ \hline
            
            \end{tabular}
    \caption{Maximum model compression achieved to produce iso-quality models. For all datasets, MEM-REC embedding tables comfortably fit in the L3 cache.}
    \label{table:iso_quality_compression}
    }
    \end{table}

\subsection{Efficiency (RQ4)} \label{sec:efficiency}

\begin{table}[htb]
    \scriptsize
    \begin{tabular}{|c|c|c|c|}
    \hline
    \textbf{LLC Size (MB)} & \textbf{14} & \textbf{28} & \textbf{56} \\ \hline
    Reduction in num cycles (latency improvement)             & 2.6x        & 3.2x        & 3.4x        \\ \hline
    Reduction in number of cache (LLC) misses (i.e. more embeddings served from caches) & 2.3x        & 6.3x        & 341x        \\ \hline
    Reduction in avg. memory BW utilization (reduce data movement b/w LLC \& DRAM) & 1.1x       & 2.2x        & 98x        \\ \hline
    \end{tabular}
    \caption{Hardware bottleneck analysis: DLRM vs MEM-REC. MEM-REC parameters used are $d=75000, d^{'}=75000, k=1, k^{'}=4, l=128$. MEMREC can perform up to $3.4\times$ faster embeddings than the DLRM baseline by reducing the LLC misses and improving bandwidth.}
    \label{table:efficiency}
    \end{table}

As previously discussed, state-of-the-art recommendation systems have large embedding tables typically of the order of 100s of GBs \citep{robe}. Random sparse accesses to these embedding table entries dominate the DLRM inference latency \citep{ke2019recnmp}. As explained in \cite{deeprecsys} commercial scale DLRM models have 10s of embedding tables each having millions of entries. In a typical commercial deployment, an inference request performs pooling over 10s to 100s of the embedding entries \citep{deeprecsys}.

To evaluate the performance of MEM-REC on a server class CPU, we configured a micro-benchmark that represents the \textit{RM2} configuration of DLRM model \citep{deeprecsys} and has a pooling factor of 120 (average pooling factor reported in \cite{deeprecsys}). We use the embedding tables which are of the same size (96GB) as that used by the Mlperf DLRM Criteo TB benchmark. Similarly,  for MEM-REC, we use the embedding table of size 46 MB that can generate an iso-quality model as discussed in Section~\ref{sec:model_compression} and with the same access profile. We ran our benchmark on a 40-Core IceLake server \footnote{Intel {IceLake} Specs, https://www.intel.in/content/www/in/en/products/platforms/details/ice-lake-sp.html} running at 2.7GHz with 204 GB/s memory bandwidth available and our results show that MEM-REC performs $3.3\times$ faster embedding on average than the DLRM baseline.

To better understand the performance implications of using different cache sizes, we also analyze MEM-REC's sensitivity to cache size by using Sniper multi-core simulator \citep{sniper}. We compare DLRM baseline and MEM-REC for different cache sizes using the total number of cycles taken, LLC miss rate and DDR bandwidth. Table ~\ref{table:efficiency} shows that as the size of last-level caches grows, MEM-REC's cache misses and memory bandwidth requirement shrink leading to up to $3.4\times$ better embedding time. 

\textbf{Takeaway 4:} MEM-REC exploits the memory access latency difference between the LLC and DRAM to improve the embedding throughput up to $3.4\times$. Furthermore, given the typical large cache sizes in modern server-class CPUs, MEM-REC can be used to reduce the data movement between LLC and DRAM by up to $98\times$ resulting in not only better throughput but also improved energy efficiency of recommendation systems.

\section{Conclusion and Future Work} \label{sec:future_work}
We presented an alternative approach to represent categorical data. We showed that a compact embedding representation obtained using MEM-REC leads to a substantial reduction in model size. Using Facebook’s DLRM framework, we demonstrated how MEM-REC can be used to achieve a memory-efficient fast recommendation pipeline without compromising the quality of the generated recommendations. In the future, we plan to study the architectural implications of the MEM-REC model on different hardware platforms. Apart from optimizing training and inference pipelines, we plan to explore how we can leverage the cache-friendliness of the MEM-REC approach to design more efficient pipelines.

\acks
{This work is partially supported by the SRC JUMP 1.0 CRISP Center, and follow on collaboration with SRC JUMP 2.0 PRISM Center, and National Science Foundation (NSF) grants \#1826967, \#1911095, \#2052809, \#2112665, \#2112167, and \#2100237.
We thank Nageen Himayat and Yash Akhauri for their expertise and assistance during this project.
}

\begin{spacing}{0.9}
\bibliography{acml23}
\end{spacing}

\end{document}